\documentclass[12pt]{article}
\usepackage{epsf}
\usepackage{amsmath,amssymb}
\usepackage{pdfpages}
\usepackage{graphicx}
\usepackage{comment}
\usepackage{tikz}
\usetikzlibrary{positioning}
\usepackage{physics}
\usepackage[driverfallback=dvipdfm]{hyperref}

\allowdisplaybreaks[1]

\begin{document}

\newcommand{\lsim}{\stackrel{<}{_\sim}}
\newcommand{\gsim}{\stackrel{>}{_\sim}}

\renewcommand{\theequation}{\thesection.\arabic{equation}}

\renewcommand{\thefootnote}{\fnsymbol{footnote}}
\setcounter{footnote}{0}

\begin{titlepage}

\def\thefootnote{\fnsymbol{footnote}}

\begin{center}

\hfill KEK-TH-2298\\
\hfill IPMU21-0012\\
\hfill February, 2021\\

\vskip .75in

{\Large \bf

  Studying squark mass spectrum through gluino decay\\[2mm]
  at 100\, TeV future hadron colliders

}

\vskip .5in

{\large So Chigusa${}^{a,b,c}$, Koichi Hamaguchi${}^{d,e}$, Takeo Moroi${}^{d,e}$,
  Atsuya Niki${}^{d}$, and Kosaku Ono${}^{d}$}

\vskip 0.25in

{\em
  $^a$Berkeley Center for Theoretical Physics, Department of Physics,\\
  University of California, Berkeley, CA 94720, USA}

\vskip 0.1in

{\em
  $^b$Theoretical Physics Group, Lawrence Berkeley National Laboratory,\\
  Berkeley, CA 94720, USA}

\vskip 0.1in

{\em
  $^c$KEK Theory Center, IPNS, KEK, Tsukuba, Ibaraki 305-0801, Japan
  }

\vskip 0.1in

{\em ${}^{d}$Department of Physics, University of Tokyo,
Tokyo 113-0033, Japan}

\vskip 0.1in

{\em ${}^{e}$Kavli IPMU (WPI), UTIAS, University of Tokyo, Kashiwa, Chiba 277-8583, Japan}

\end{center}
\vskip .5in

\begin{abstract}

  We study the prospect of determining the decay properties of the gluino
  in the supersymmetric (SUSY) standard model at a 100 TeV future hadron
  collider.  We consider the case where the neutral Wino is the
  lightest superparticle. In this case, the long-lived charged Wino can be
  used to eliminate standard model backgrounds, which enables us to
  study the details of superparticles.  We show that, based on the
  analysis of the numbers of high $p_T$ leptons, boosted $W$-jets, and
  $b$-tagged jets, we may determine the gaugino species and the quark
  flavors in the gluino decay.  With such determinations, we can
  obtain information about the mass spectrum of squarks even if
  squarks are out of the kinematical reach.

\end{abstract}

\end{titlepage}

\renewcommand{\thepage}{\arabic{page}}
\setcounter{page}{1}
\renewcommand{\thefootnote}{\#\arabic{footnote}}
\setcounter{footnote}{0}

\section{Introduction}
\label{sec:intro}
\setcounter{equation}{0}

Collider experiments at the energy frontier are important in
understanding the properties of elementary particles.  In the last
decade, the Large Hadron Collider (LHC) has not only discovered the
Higgs boson~\cite{Aad:2012tfa,Chatrchyan:2012ufa} but also has
revealed its properties.  The results from the LHC experiment (as well
as those from other experiments) have been essential to confirm the
validity of the standard model (SM) as the effective theory for the
energy scale below the TeV scale.  Despite the success of the SM,
however, it is widely believed that the SM is not the ultimate theory
and that there should show up physics beyond the SM (BSM).  This is
because there are still mysteries that cannot be explained in the
framework of the SM; for example, from the particle physics point of
view, the charge quantization (which is naturally explained in the
grand unified theory (GUT)) cannot be explained in the standard model,
and from cosmology point of view, the origin of dark matter is not
understood.

One of the tasks of the future energy frontier experiments is to find
signals of BSM physics and to study its properties.  In the next
decade, the LHC Run-3/HL-LHC will play such a role and will try to discover
BSM particles.  However, the searches are limited by the collider
energy and the discovery is impossible if the BSM particles are out of
the kinematical reach.  For example, in the supersymmetric (SUSY) SM
with the Wino lightest superparticle (LSP), the thermal Wino can
become dark matter if its mass is about $2.9\, {\rm TeV}$ and hence
such a model is well motivated.  However, such a heavy Wino is out of
the reach of the LHC experiment~\cite{Saito:2019rtg}.  For the
discovery and the study of the BSM particles, we may need a collider
with the center of mass energy much higher than the LHC.

Recently, the possibilities of such high energy colliders have been
discussed.  In this paper, we consider circular $pp$ collider with the
center of mass energy of about $100\, {\rm TeV}$, i.e., future
circular collider (or dubbed as FCC-hh).  FCC-hh is a prominent
candidate for a future energy frontier
experiment~\cite{Golling:2016gvc}.

Here, we discuss the prospect of studying the properties of
superparticles in SUSY SM at the FCC-hh.  We will pay particular
attention to the so-called pure gravity mediation model of SUSY
breaking \cite{Ibe:2006de, Ibe:2011aa, ArkaniHamed:2012gw} based on
anomaly mediation \cite{Randall:1998uk, Giudice:1998xp}; such a model
naturally results in the Wino LSP (possibly with the Wino mass of
$\sim 2.9\, {\rm TeV}$) so that the Wino can be a viable dark matter
candidate.  In addition, with the introduction of SUSY particles, the
gauge coupling unification at the GUT scale of about $10^{16}\, {\rm
  GeV}$ is suggested.  Thus, even though the signal of SUSY has not
been discovered yet, it is still an attractive candidate for the BSM
physics.  It is important to understand what we can learn about such a
model at the FCC-hh.  Indeed, there have been efforts to investigate
the potential of the FCC-hh for the study of the pure gravity
mediation SUSY model.  It has been discussed that the FCC-hh will
discover the Wino LSP using the fact that the decay length of the
charged Wino may become macroscopic \cite{Saito:2019rtg}.  It has also
been pointed out that the mass spectrum of the gauginos and the
lifetime of the charged Wino can be studied at the FCC-hh after the
discovery of the Wino LSP~\cite{Asai:2019wst, Chigusa:2019zae}.

In this paper, a possibility to study the decay properties of the
gluino $\tilde{g}$ at the FCC-hh is discussed.  (For related studies
at the LHC, see \cite{Sato:2012xf, Sato:2013bta}.)  In the pure
gravity mediation model, the gluino decays into a quark anti-quark
pair and a Wino $\tilde{W}$ or a Bino $\tilde{B}$.  We will show that,
by analyzing the numbers of high $p_T$ leptons, boosted $W$-jets, and
$b$-tagged jets in the gluino production events, the gaugino species
and the quark flavors in the gluino decay may be understood.  As the
partial branching ratios of the gluino are model dependent and
sensitive to the squark masses, the detailed study of the gluino decay
may give information about the mass spectrum of squarks which may be
out of the kinematical reach of the FCC-hh.

The organization of this paper is as follows.  In Section
\ref{sec:model}, we introduce our representative model based on which
we perform the Monte Carlo (MC) analysis.  In Section
\ref{sec:analysis}, we explain the detail of our MC analysis and show
the numerical results.  In Section \ref{sec:implication}, we discuss
implications of the determination of the partial branching ratios of
the gluino at the FCC-hh.  The results are summarized in Section
\ref{sec:summary}.

\section{Model}
\label{sec:model}
\setcounter{equation}{0}

We first introduce the SUSY model of our interest.  As we have
mentioned, we consider the pure gravity mediation model in which the
gaugino masses are from the effect of anomaly mediation while the SUSY
breaking scalar mass squared parameters originate from the
supergravity effect.  In such a framework, the scalar masses and the
Higgsino mass $\mu$ are of the order of the gravitino mass $m_{3/2}$
while the gaugino masses are one-loop suppressed relative to the
scalar masses.  The model has several phenomenological advantages.
First, the SM-like Higgs mass can be pushed up to the observed value
of $\sim 125\, {\rm GeV}$ \cite{Zyla:2020zbs} by radiative corrections
\cite{Okada:1990vk, Ellis:1990nz, Haber:1990aw}.  Second, the heavy
sfermion masses suppress the CP and flavor violating processes
mediated by SUSY particles in the loop, which significantly relaxes
the SUSY CP and flavor problems.  (See, however, \cite{Moroi:2013sfa,
  McKeen:2013dma}.)  Furthermore, cosmologically, the neutral Wino can
be a viable candidate of dark matter; in particular, if the Wino mass
is $\sim 2.9\, {\rm TeV}$, the thermal relic abundance of Wino becomes
consistent with the present dark matter density \cite{Hisano:2006nn}.
Motivated by these, in the following, we consider the pure gravity
mediation model with gaugino masses of $O(1)\, {\rm TeV}$ and the
gravitino mass of $O(10-100)\ {\rm TeV}$.

In such a model, at the FCC-hh with the center-of-mass energy of $\sim
100\, {\rm TeV}$, the gauginos are the primary targets while the
sfermions may be hardly produced.  Hereafter, we consider the case
where the gauginos are accessible with the FCC-hh while the sfermions
are out of the kinematical reach.  For our numerical analysis in the
next section, we adopt the mass spectrum of the gauginos suggested by
the model of pure gravity mediation; the sample points adopted in our
analysis are summarized in Table~\ref{table:samplept}, in which the
Bino, Wino, and gluino masses (denoted as $m_{\tilde{B}}$,
$m_{\tilde{W}}$, and $m_{\tilde{g}}$, respectively), as well as the
gluino pair production cross section for $\sqrt{s}=100\ {\rm TeV}$,
are shown.\footnote
{In our analysis, we use the leading order estimation of the gluino
  production cross section.  Including the next-to-leading
  order effects, the cross section increases by $O(10)\ \%$
  \cite{Golling:2016gvc}.}
We assume that the Wino is the LSP with a mass of 2.9 TeV.
For more details of the sample points, see Ref.~\cite{Asai:2019wst}.

\begin{table}[t]
  \begin{center}
    \begin{tabular}{c|cc}
      \hline\hline
      & Point 1 & Point 2
      \\
      \hline
      $m_{\tilde{B}}$ [GeV] & $3660$ & $4060$
      \\
      $m_{\tilde{W}}$ [GeV] & $2900$ & $2900$
      \\
      $m_{\tilde{g}}$ [GeV] & $6000$ & $7000$
      \\
      $\sigma (pp\rightarrow \tilde{g}\tilde{g})$ [fb]
      & $7.9$ & $2.7$
      \\
      \hline\hline
    \end{tabular}
    \caption{Gaugino masses and gluino pair production cross
      section (for the center-of-mass energy of $100\ {\rm TeV}$) for the
      Sample Points 1 and 2.}
    \label{table:samplept}
  \end{center}
\end{table}

In discussing the collider phenomenology of the model of our interest,
one remarkable feature is the property of the Wino LSP.  Because the
Wino is in the adjoint representation of $SU(2)_L$, there exist
neutral and charged Winos, $\tilde{W}^0$ and $\tilde{W}^\pm$,
respectively.  Without the effects of the electroweak symmetry
breaking the masses of neutral and charged Winos are degenerate.  The
electroweak symmetry breaking generates a small mass gap between the
neutral and charged Winos through a loop effect and the charged Winos
dominantly decay as $\tilde{W}^\pm\rightarrow\tilde{W}^0\pi^\pm$.
Because of the high mass degeneracy, the lifetime of the charged Wino
becomes significantly long; the most precise calculation gives
\cite{Ibe:2012sx}
\begin{align}
  \label{eq:lifetime_wino}
  c\tau\simeq 5.75\, {\rm cm},
\end{align}
where $\tau$ is the lifetime while $c$ is the speed of light.
When $\mu\gg m_{\tilde{W}}\gg m_W$, the lifetime is almost independent of
the Wino mass.
Then,
once produced at the FCC, the charged Wino may fly a macroscopic
length of $O(1-10)\, {\rm cm}$.  Such a long-lived charged Wino may be
identified if it goes through several layers of the inner pixel
detector of the FCC-hh, and it can be regarded as a characteristic
feature of the SUSY signal.  As we will discuss in the next section,
the long-lived charged Wino can be used to remove the SM background.

The subject of this paper is to discuss the possibility of studying
the decay properties of the gluino at the FCC-hh.  Here, for simplicity,
we consider the case where the flavor violating decay of the gluino is
negligible.  Then, the partial decay rates of the gluino are given
by~\cite{BARBIERI198815,Gambino:2005eh,Sato:2013bta}\footnote
{Here, we neglect the renormalization group effect on the partial
  decay rates.  At the one-loop level, the effects of the strong gauge
  coupling constant, which are the most important, are universal to
  all the final-state quark flavors and does not affect our later
  discussion about the study of the ratios of squark masses.  In
  addition, the effects of the top Yukawa interaction, which is
  relevant only for the processes with third generation quarks in the
  final state, is of $O(1)\ \%$ for the parameter region we
  consider. (See, for example, \cite{Sato:2013bta}.)  We also neglect
  the effects of the left-right mixing and the two-body decay
  $\tilde{g}\rightarrow\tilde{B}g$ (with $g$ being gluon), which are
  unimportant for the case of our study \cite{Gambino:2005eh}.}
\begin{align}
  \Gamma (\tilde{g}\rightarrow \tilde{B}u_{Li}\bar{u}_{Li}) =
  \Gamma (\tilde{g}\rightarrow \tilde{B}d_{Li}\bar{d}_{Li}) =   &\,
  \frac{1}{1536\pi^3}\frac{g_s^2 g'^2}{36}\frac{m_{\tilde{g}}^5}{m_{\tilde{Q}_i}^4} f\left(\frac{m_{\tilde{B}}}{m_{\tilde{g}}}\right)\,,
  \label{eq:gluino_decay_Bqq}
  \\
  \Gamma (\tilde{g}\rightarrow \tilde{B}u_{Ri}\bar{u}_{Ri}) = &\,
  \frac{1}{1536\pi^3}\frac{4g_s^2 g'^2}{9}\frac{m_{\tilde{g}}^5}{m_{\tilde{u}_i}^4} f\left(\frac{m_{\tilde{B}}}{m_{\tilde{g}}}\right)\,,
  \\
  \Gamma (\tilde{g}\rightarrow \tilde{B}d_{Ri}\bar{d}_{Ri}) = &\,
  \frac{1}{1536\pi^3}\frac{g_s^2 g'^2}{9}\frac{m_{\tilde{g}}^5}{m_{\tilde{d}_i}^4} f\left(\frac{m_{\tilde{B}}}{m_{\tilde{g}}}\right)\,,
  \\
  \Gamma (\tilde{g}\rightarrow \tilde{W}^0u_{Li}\bar{u}_{Li}) =
  \Gamma (\tilde{g}\rightarrow \tilde{W}^0d_{Li}\bar{d}_{Li}) = &\,
  \frac{1}{1536\pi^3}\frac{g_s^2 g^2}{4}\frac{m_{\tilde{g}}^5}{m_{\tilde{Q}_i}^4} f\left(\frac{m_{\tilde{W}}}{m_{\tilde{g}}}\right)\,,
  \\
  \Gamma (\tilde{g}\rightarrow \tilde{W}^-u_{Li}\bar{d}_{Li}) =
  \Gamma (\tilde{g}\rightarrow \tilde{W}^+d_{Li}\bar{u}_{Li}) = &\,
  \frac{2}{1536\pi^3}\frac{g_s^2 g^2}{4}\frac{m_{\tilde{g}}^5}{m_{\tilde{Q}_i}^4} f\left(\frac{m_{\tilde{W}}}{m_{\tilde{g}}}\right)\,,
  \label{eq:gluino_decay_Wpm}
\end{align}
where $i=1-3$
denotes generation index, and
\begin{align}
  f(x) = &\,
  1 - 8|x|^2 -12|x|^4 \ln |x|^2 + 8|x|^6  - |x|^8
  \nonumber \\ &\,
  + 2(1+9|x|^2 + 6|x|^2 \ln |x|^2 - 9|x|^4 + 6|x|^4\ln |x|^2 - |x|^6)
  {\rm Re}(x).
\end{align}
Because we are interested in the case where the gluino mass is much
larger than the quark masses, we neglect the quark masses.  As one can
see, the partial decay rates are sensitive to the mass spectrum of
squarks and are highly model dependent.  Thus, with the detailed
studies of branching ratios of the gluino, we can obtain information
about the mass spectrum of squarks.

In the sample points we adopted, the Bino is unstable and decays into
a charged or neutral Wino.
In the limit of $\mu\gg m_{\tilde{B}/\tilde{W}}\gg m_W$, the partial decay rates
of the dominant decay processes of the Bino are insensitive to the
sfermion masses, and the Bino dominantly decays into $\tilde{W}^\pm
W^\mp$ or $\tilde{W}^0h$ with branching fractions of
\begin{align}
  {\rm Br} (\tilde{B}\rightarrow \tilde{W}^+ W^-) =
  {\rm Br} (\tilde{B}\rightarrow \tilde{W}^- W^+) \simeq
  {\rm Br} (\tilde{B}\rightarrow \tilde{W}^0h) \simeq
  \frac{1}{3}.
\end{align}

\section{Analysis}
\label{sec:analysis}
\setcounter{equation}{0}

\subsection{Setup}

In this section, we discuss the measurement of the branching ratios of
the gluino decay at a 100 TeV collider. We consider the pair production of
the gluino:
\begin{align}
  pp&\to \tilde{g}\tilde{g},
\end{align}
followed by the decay of each gluino with a charged Wino in the final state:
\begin{align}
  \tilde{g}\to
  \begin{cases}
     \tilde{B} q \bar{q},\, \mbox{with}\,
     \tilde{B}\to \tilde{W}^\pm W^\mp,
     \\[1mm]
     \tilde{W}^\pm q\bar{q}.
  \end{cases}
\end{align}
As discussed in \cite{Saito:2019rtg}, for the sample points we have
adopted, the Wino is within the discovery reach of the FCC-hh using a
disappearing track signature.  After the discovery of the gauginos,
all the gaugino masses can be measured at the FCC-hh
\cite{Asai:2019wst}.  In addition, the lifetime of charged Wino can be
determined by analyzing the distribution of the flight length
\cite{Chigusa:2019zae}.  Thus, in the following analysis, we assume
that we can use the information about the gaugino masses and the Wino
lifetime and discuss how and how well we can determine the branching
ratios of the gluino decay processes.  In particular, the information
about the gluino mass is essential to predict the production cross
section of the gluino pair, while the Wino lifetime is necessary to
determine the detection rate of the long-lived charged Wino at the
inner pixel detector (see discussion below).  In principle, the cross
section for the process $pp\to\tilde{g}\tilde{g}$, as well as the
survival detection probability of the charged Wino, can be
theoretically calculated once the gaugino masses and the Wino lifetime
are known.  In our analysis, we assume that reliable calculations of
these quantities are possible at the time of the FCC-hh experiment.
We neglect systematic uncertainties in our MC analysis and comment on
them at the end of this section.

As shown in Eqs.\ \eqref{eq:gluino_decay_Bqq} --
\eqref{eq:gluino_decay_Wpm}, ${\rm Br}(\tilde{g}\to
\tilde{W}q\bar{q})$ becomes large (small) when the left-handed squarks
are light (heavy) compared to the right-handed ones. Moreover, when
the third generation squarks are light (heavy), the branching ratios
into the third generation quarks become enhanced (suppressed).
Motivated by these features, we study how well we can constrain (i)
the probability that the gluino decays into a Bino, not a Wino (which we call
$x$), and (ii) the probability that the gluino decays into the third
generation quark anti-quark pair rather than the first or second
generation one (which we call $y$):
\begin{align}
\sum_q {\rm Br}(\tilde{g}\to \tilde{B} q\bar{q}) &= x,
\label{Eq:x}
\\
\quad \sum_{q,q'} {\rm Br}(\tilde{g}\to \tilde{W} q\bar{q}') &= 1-x,
\\
\sum_{q,q'=t,b} \left[
{\rm Br}(\tilde{g}\to \tilde{B} q\bar{q}) +
{\rm Br}(\tilde{g}\to \tilde{W} q\bar{q}')
\right] &= y,
\\
\sum_{q,q'=u,d,c,s} \left[
{\rm Br}(\tilde{g}\to \tilde{B} q\bar{q}) +
{\rm Br}(\tilde{g}\to \tilde{W} q\bar{q}')
\right] &= 1-y.
\label{Eq:1-y}
\end{align}
Thus, in our analysis, a model point is characterized by a set of
$(x,y)$ as well as gaugino masses.
More concretely, for our numerical calculation, the gluino
branching ratios are set as follows:
\begin{align}
{\rm Br}(\tilde{g}\to \tilde{B} q\bar{q})
&=
\begin{cases}
\frac{1}{4}x(1-y) & (q=u,d,c,s)
\\
\frac{1}{2}x y & (q=t,b)
\end{cases},
\label{G2B}
\\
{\rm Br}(\tilde{g}\to \tilde{W}^0 q\bar{q})
&=
\begin{cases}
\frac{1}{12}(1-x)(1-y) & (q=u,d,c,s)
\\
\frac{1}{6}(1-x) y & (q=t,b)
\end{cases},
\\
{\rm Br}(\tilde{g}\to \tilde{W}^- q\bar{q}')
=
{\rm Br}(\tilde{g}\to \tilde{W}^+ q'\bar{q})
&=
\begin{cases}
\frac{1}{6}(1-x)(1-y) & (q\bar{q}'=u\bar{d},c\bar{s})
\\
\frac{1}{3}(1-x) y & (q\bar{q}'=t\bar{b})
\end{cases}.
\label{G2Wpm}
\end{align}
Flavor violating decay processes of the gluino are assumed to be
negligible.  
For larger (smaller) $x$, the numbers of leptons and boosted $W$-jets
increase (decrease) because they are produced by the decay of the Bino.
(Notice that high $p_T$ leptons are produced by the decay of $W$
bosons from the Bino decays.)  In addition, larger $y$ is expected to enhance
the number of $b$-tagged jets.
The parameter space in our analysis is thus $0\le x,y\le 1$ and
we study how well we can determine
the $x$ and $y$ parameters at the FCC-hh using these features in the
following.

We comment that, in general, the partial branching ratios cannot be
determined just by $x$ and $y$, and that Eqs.\ \eqref{G2B} --
\eqref{G2Wpm} are examples which realize Eqs.\ \eqref{Eq:x} --
\eqref{Eq:1-y}.  As we will see below, the accuracy of the $x$
determination is insensitive to the quark flavors from the gluino
decay (i.e., the choice of $y$) while that of the $y$ determination
does not depend so much on the gaugino species (i.e., the choice of
$x$).  Thus, we expect that our main conclusions are not significantly
altered by the detail of the partial branching ratios for a given set
of $(x,y)$.

\subsection{Method}
In the sample points we take, the gluino is within the kinematical reach of
the FCC, and the gluino pair is produced as
$pp\rightarrow\tilde{g}\tilde{g}$.  If signals are selected only using
a missing $E_T$ cut, a significant amount of SM backgrounds are
expected.  In order to eliminate the SM backgrounds, we use the fact
that the signal on inner pixel detectors given by charged Winos can be
used to identify the SUSY events.  As we have mentioned, the decay
length of the charged Wino can be as long as $\sim 10\ {\rm cm}$.  Such a
long-lived charged Wino hits several layers of the inner pixel
detector and, after the decay, it does not leave any
energetic activity in outer
detectors.  Then, the long-lived charged Wino is regarded as a short
high $p_T$ track, which is hardly mimicked by SM events.  Thus, by
requiring long-lived charged Wino tracks, a significant reduction of the
SM backgrounds is expected.

In our analysis, we impose the following requirements on the signal
events:
\begin{enumerate}
\item The missing transverse energy $E_T$ is larger than $1\ \rm{TeV}$.
\item Each gluino has charged Wino in its decay chain; each charged
  Wino is assumed to be identified by the inner pixel detector with
  imposing the following requirements 3 and 4.
\item The pseudorapidities $\eta$ of both charged Winos are smaller
  than $1.5$.
\item The transverse flight lengths $L_T$ of both charged Winos should
  be longer than $10\ \rm{cm}$.  (Here, we assume that the transverse
  distance to the fourth layer of the pixel detector is $10\ {\rm cm}$
  so that each charged Wino goes through four layers of the pixel
  detector.)
\end{enumerate}
We assume that the SM backgrounds become negligible after imposing
these requirements~\cite{Saito:2019rtg, Asai:2019wst}.

By using the information from the signal events, we can determine the
partial decay rate of the gluino as we discuss in detail in the following.
For each event, we count the numbers of leptons ($e^\pm$ and
$\mu^\pm$), boosted $W$-jets, and $b$-tagged jets:
\begin{itemize}
\item We use leptons and $b$-tagged jets whose transverse momenta are
  larger than $200\ \rm{GeV}$.
\item We define the ``boosted $W$-jets'' as jets with mass $60\ {\rm
  GeV} < m_{jet} < 100\ {\rm GeV}$, the ratio of $N$-subjettiness
  \cite{Thaler:2010tr} $\tau_2/\tau_1 < 0.3$
  \cite{Khachatryan:2014vla}, and $p_T>200\ {\rm GeV}$.
\end{itemize}

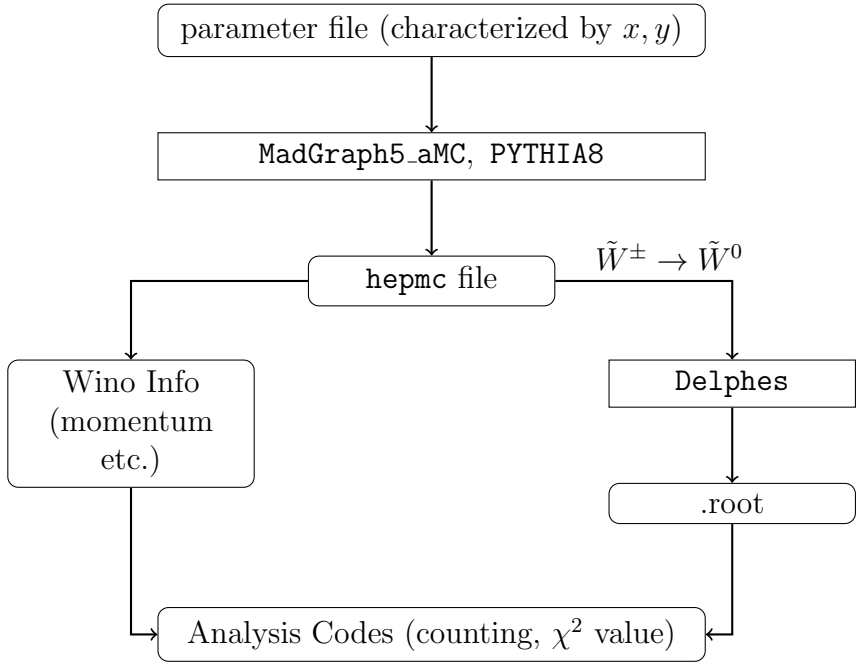
\begin{figure}
  \begin{center}
    \begin{tikzpicture}
      \tikzset{round1/.style={rectangle, draw, text centered, text width=7cm, rounded corners, minimum height=0.5cm}};
      \tikzset{round2/.style={rectangle, draw, text centered, text width=3cm, rounded corners, minimum height=0.5cm}};
      \tikzset{rect1/.style={rectangle, draw, text centered, text width=7cm, minimum height=0.5cm}};
      \tikzset{rect2/.style={rectangle, draw, text centered, text width=3cm, minimum height=0.5cm}};
      \node[round1](parameter){parameter file (characterized by $x,y$)};
      \node[rect1,below=of parameter](mad){${\tt MadGraph5\_aMC},\ {\tt PYTHIA8}$};
      \draw[->,thick] (parameter)--(mad);
      \node[round2,below=of mad](hepmc){{\tt hepmc} file};
      \draw[->,thick] (mad)--(hepmc);
      \node[round2,below left=1.0 of hepmc](wino){Wino Info \\(momentum etc.)};
      \node[rect2,below right=1.0 of hepmc](delphes){${\tt Delphes}$};
      \draw[->,thick] (hepmc)-|(wino);
      \draw[->,thick] (hepmc)node[above,xshift=90]{$\tilde{W}^\pm\rightarrow\tilde{W}^0$}-|(delphes);
      \node[round2,below=of delphes](root){.root};
      \draw[->,thick] (delphes)--(root);
      \node[round1,below=4 of hepmc](analysis){Analysis Codes (counting,\ $\chi^2$ value)};
      \draw[->,thick] (wino)|-(analysis);
      \draw[->,thick] (root)|-(analysis);
    \end{tikzpicture}
  \end{center}
  \caption{The flowchart of MC simulation. Rectangular means simulator
    and rounded corner rectangular means data file and analysis
    codes.}
  \label{MC_simulation}
\end{figure}

In order to see how well we can constrain the $x$ and $y$ parameters, we
perform an MC analysis.  The flowchart of our MC simulation is shown in
Fig.~\ref{MC_simulation}. We use ${\tt
  MadGraph5\_aMC@NLO\ 2.7.2}$\ \cite{Alwall:2011uj,Alwall:2014hca} to
generate $pp \rightarrow \tilde{g}\tilde{g}$ events. Decay and
hadronization processes are simulated by using ${\tt PYTHIA8}$
\cite{Sjostrand:2014zea}. Detector simulation is done by ${\tt Delphes\ 3.4.2}$
\cite{deFavereau:2013fsa} using ${\tt FCChh.tcl}$ card.

The disappearing track of charged Winos cannot be simulated by {\tt
  Delphes} by default; charged Winos are treated in a similar way as
other charged particles in {\tt Delphes}.  In order to simulate the
decay and the detection of charged Winos, we calculate the
flight-length distribution of each charged Wino using the information
provided by {\tt hepmc} file (which is the output of ${\tt PYTHIA8}$),
while each charged Wino is treated as non-detectable neutral particles
in {\tt Delphes}.  (In the {\tt Delphes} simulation, we use the {\tt
  hepmc} file in which the particle ID number of charged Wino is
changed to that of neutral Wino.)

Combining the output of ${\tt Delphes}$ simulation and the
flight-length distributions of charged Winos, we calculate the
distributions of the number of leptons, boosted $W$-jets, and $b$-tagged jets.
The analysis is performed by using ${\tt ROOT\ 6.18}$ \cite{Brun:1997pa}.

Now we explain how we study the prospects of constraining $x$ and $y$
parameters at the FCC-hh.  In our analysis, the parameter space is
discretized as $x,y \in [0.0,0.1,0.2,...,1.0]$. We determine the
partial branching ratios of the gluino for given $x$ and $y$ (see
Eqs.\ \eqref{G2B} -- \eqref{G2Wpm}).  Then, for each set of $(x,y)$,
we simulate $pp \rightarrow \tilde{g}\tilde{g}$ process and choose
signal events satisfying the requirements 1 -- 4 introduced before.
An example of the cut flow, taking $x=0.5$ and $y=0.5$, is shown in
Table~\ref{table:selection}.  Then, using the signal events passing
the cuts, we calculate the expected numbers of leptons, boosted
$W$-jets, and $b$-tagged jets.  With the distributions of these
numbers on the $x$ vs.\ $y$ plane, we determine the accuracy of the
determinations of $x$ and $y$ parameters for given values of the
luminosity.

\begin{table}[t]
  \centering
  \begin{tabular}{c|r}
    \hline
    \hline
    condition&number of events\\
    \hline
    total events& $100000$ \\
    two $\tilde{W}^{\pm}$ & $44341$ \\
    $\eta\leq 1.5$& $25300$\\
    MET: $E_T\geq 1\ {\rm TeV}$& $19992$\\
    $L_T\geq 10\ {\rm cm}$& $536$\\
    \hline
    \hline
  \end{tabular}
  \caption{The cut flow for the sample point 1 with $x=0.5,\ y=0.5$,
    using 100000 events of gluino pair production.}
  \label{table:selection}
\end{table}

Here, we use the following classifications of the events:
\begin{itemize}
\item The total number of leptons ($e^\pm$ and $\mu^\pm$) is zero
  (L0) or non-zero (L1).
\item The number of boosted $W$-jet is zero (W0) or non-zero (W1).
\item The number of $b$-tagged jet is less than 2 (B01) or
  2 or larger (B2).
\end{itemize}
Based on the above classifications, we define eight signal regions
characterized by (L$\ell$, W$w$, B$b$), where $\ell=0$ or $1$, $w=0$
or $1$, and $b=01$ or $2$; all the signal events passing the
requirements 1 -- 4 are classified into one of eight signal
regions.  For a given luminosity, we calculate the expected numbers of
events falling into eight signal regions (denoted as $N_i$ with $i=1$
-- $8$).  In our analysis, $N_i$ is calculated as
\begin{align}
  N_i =
  \frac{{\cal L} \sigma_{pp\rightarrow \tilde{g}\tilde{g}}}{N_{\rm MC}}
  \sum_{A=1}^{N_{\rm MC}}
  \delta_{A}^{(\rm {cuts})}
  \delta_{({\rm L}\ell,{\rm W}w,{\rm B}b)_A,({\rm L}\ell,{\rm W}w,{\rm B}b)_i}
  P_A^{(1)} P_A^{(2)},
\end{align}
where the summation is over all the event samples generated in the MC
analysis, and $N_{\rm MC}$ is the total number of event samples (which
is taken to be 100000 in our analysis).  Here, $\delta_{A}^{(\rm
  {cuts})}$ is $1$ ($0$) if $A$-th event satisfies (does not satisfy)
the kinematical requirements 1 -- 3, while $\delta_{({\rm L}\ell,{\rm
    W}w,{\rm B}b)_A,({\rm L}\ell,{\rm W}w,{\rm B}b)_i}$ is $1$ ($0$)
if $A$-th event falls (does not fall) into $i$-th signal region.  In
addition, $P_A^{(1)}$ ($P_A^{(2)}$), which takes care of the
requirement 4, is the probability that the transverse flight length of
the first (second) charged Wino produced in $A$-th event sample is
longer than $L_0=10\ {\rm cm}$.\footnote
{The probability that the transverse flight length of a charged Wino
  is longer than $L_0$ is
\begin{align*}
  P = \exp\qty(\frac{-L_0}{c\tau \beta \gamma}),
\end{align*}
where $c\beta$ is the velocity of the Wino and
$\gamma\equiv(1-\beta^2)^{-1/2}$.}

We calculate the expected numbers of events in eight signal regions
for the $11\times 11$ different choices of $(x,y)$; the result is
denoted as $N_i^{(x,y)}$.  Once the set of $N_i^{(x,y)}$ is obtained,
we perform the $\chi^2$ analysis to estimate the expected accuracy of
the determination of $(x,y)$.  The difference of the $\chi^2$ variable
between one model point with $(x_0,y_0)$, called reference point,
and another with $(x,y)$, called trial point, is given by
\begin{equation}
  \Delta\chi^2 =
  \sum_i \frac{\qty(N_i^{(x,y)}-N_i^{(x_0,y_0)})^2}{N_i^{(x_0,y_0)}}.
  \label{eq:chi2}
\end{equation}
In our analysis, this value follows $\chi^2$ distribution with two
degrees of freedom.

\subsection{Numerical results}
\label{subsec:results}

\begin{figure}[t]
  \centering
  \includegraphics[width=15.0cm]{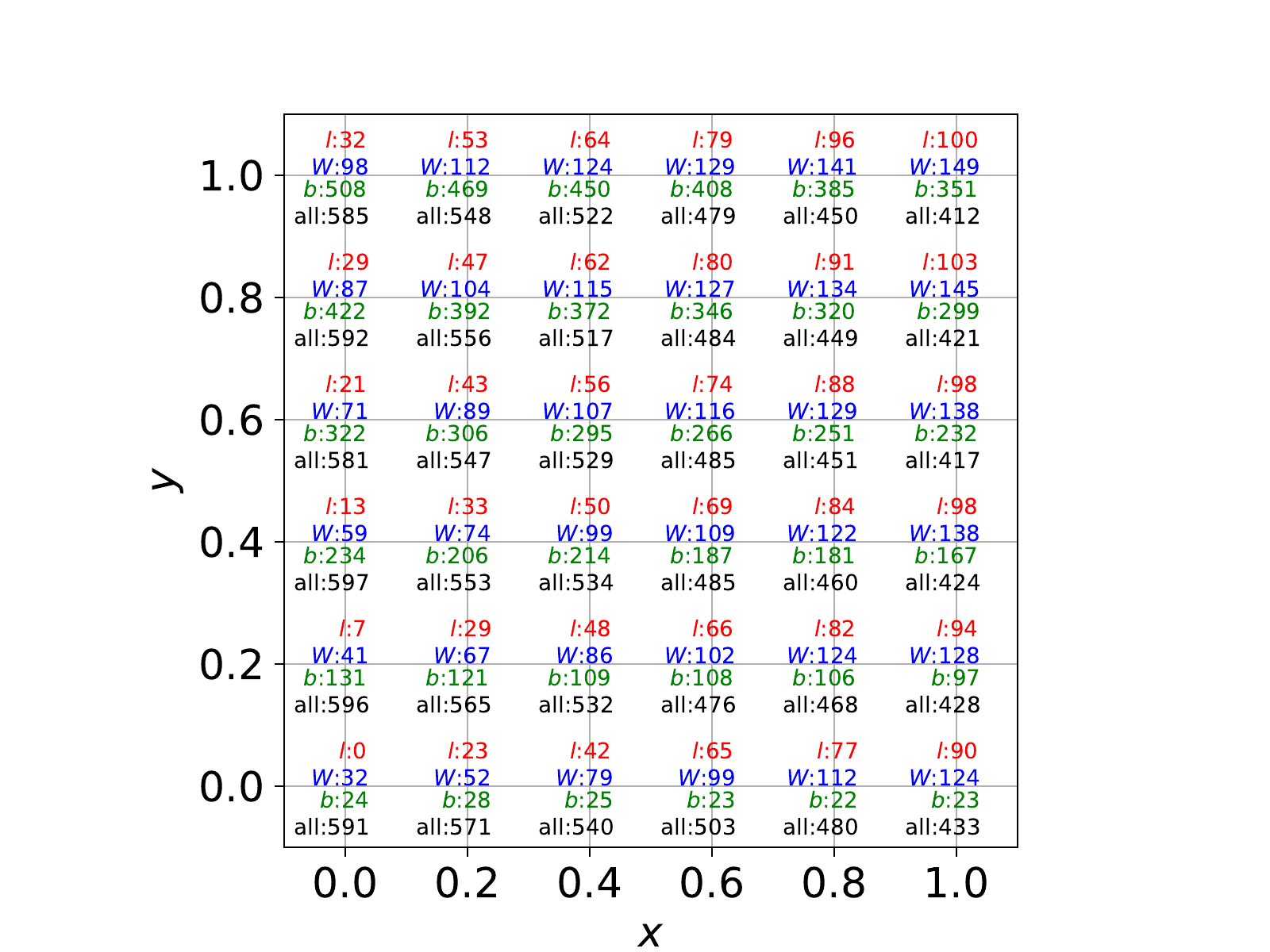}
  \caption{The numbers of events categorized in L1 (red), W1 (blue),
    and B2 (green), as well as the total number of the signal events
    (black) for the sample point 1, taking the integrated luminosity
    of $10\, {\rm ab}^{-1}$.}
  \label{fig:eventnumbers}
\end{figure}

Now, we show our numerical results.  Before discussing the expected
accuracies in the $x$ and $y$ determination, let us see how the
numbers of leptons, boosted $W$-jets, and $b$-tagged jets depend on
$x$ and $y$.  In Fig.~\ref{fig:eventnumbers}, red, blue, and green
numbers are the numbers of events categorized in L1 (with any numbers
of $W$-jet and $b$-tagged jet), W1, and B2, respectively, for some
choices of $(x,y)$.  In addition, we also show the total number of
signal events in black.  We can see that the total number of signal
events decreases as $x$ increases.  This is because, for a larger
value of $x$, the average number of final state particles becomes
larger and the averaged velocity of the charged Wino becomes smaller,
resulting in the suppression of the survival probability of
$\tilde{W}^\pm$.  As expected, the numbers of leptons and boosted
$W$-jets increase as the $x$ parameter becomes larger; this is because
the leptons and boosted $W$-jets originate from the decay process
$\tilde{B}\rightarrow\tilde{W}^\pm W^\mp$.  They also depend on $y$
because of the $W$ boson from the top quark decay.  The number of
$b$-tagged jets shows a significant dependence on $y$.  The number of
$b$-tagged jets also depends slightly on $x$, which is mainly due to
the $x$-dependence of the total number of signal events.  Thus, we can
expect that the $x$ and $y$ parameters can be constrained with the
procedure explained in the previous subsection.

\begin{figure}
  \centering
  \includegraphics[width=0.48\hsize]{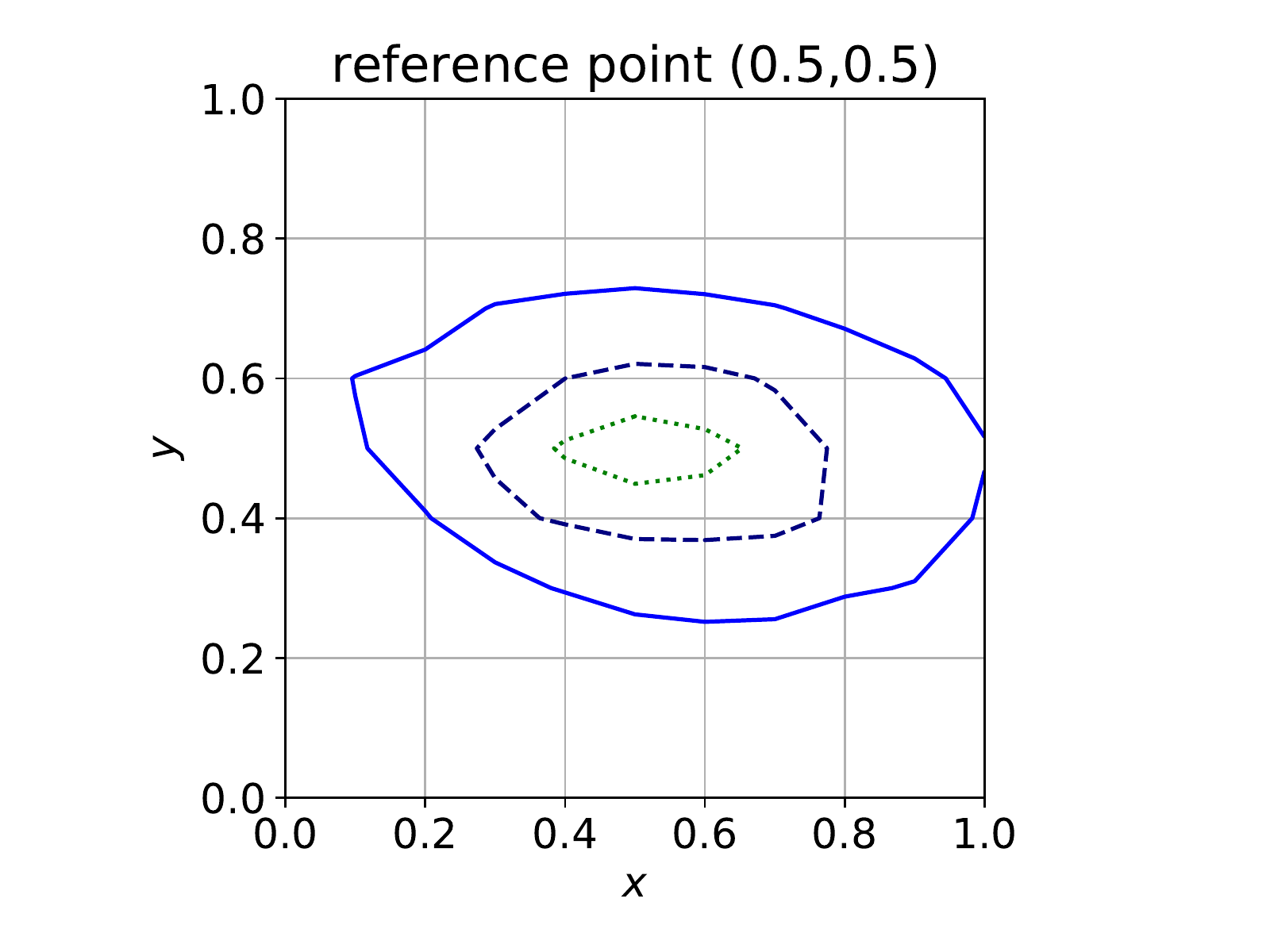}
  \includegraphics[width=0.48\hsize]{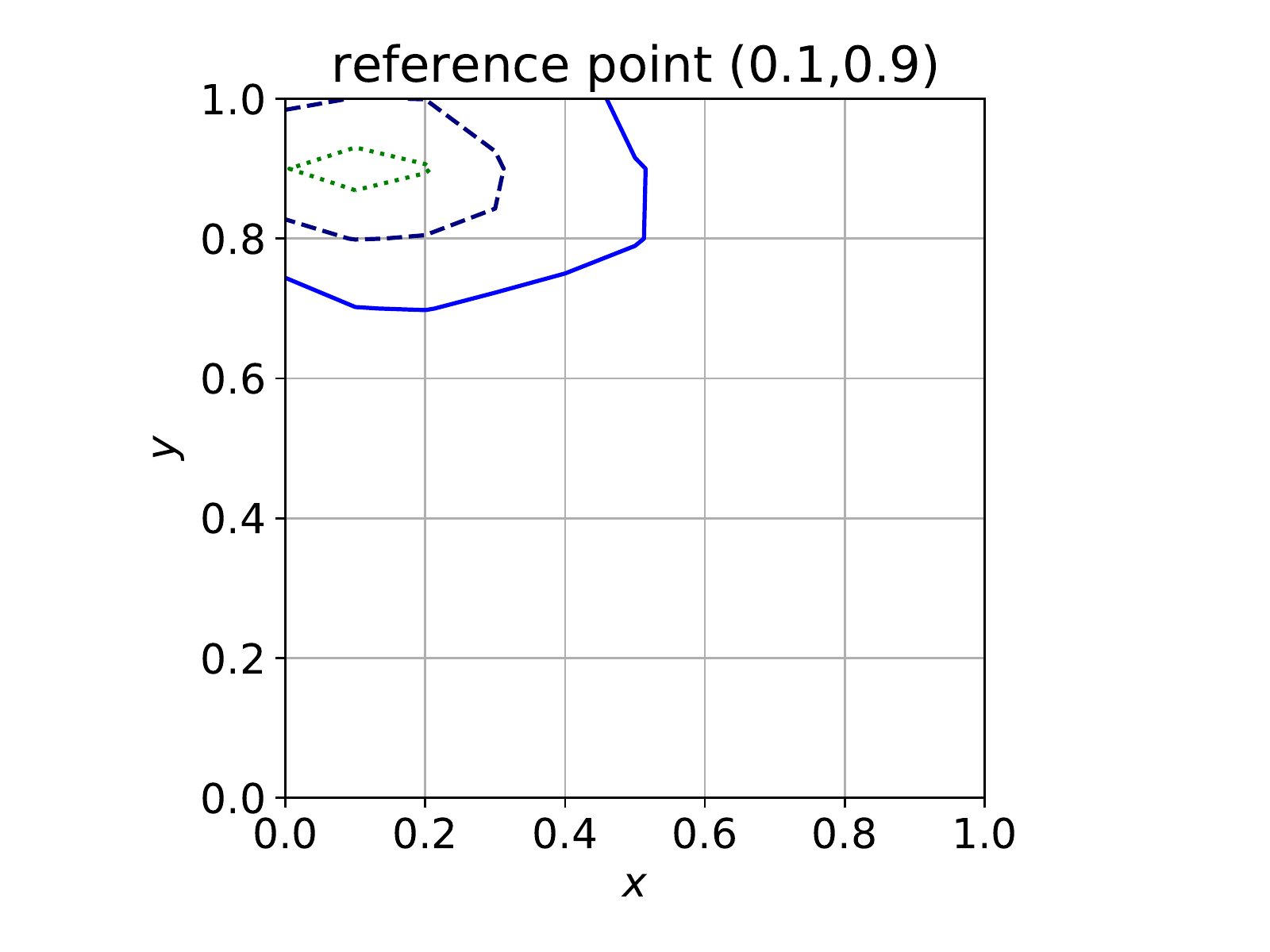}
  \includegraphics[width=0.48\hsize]{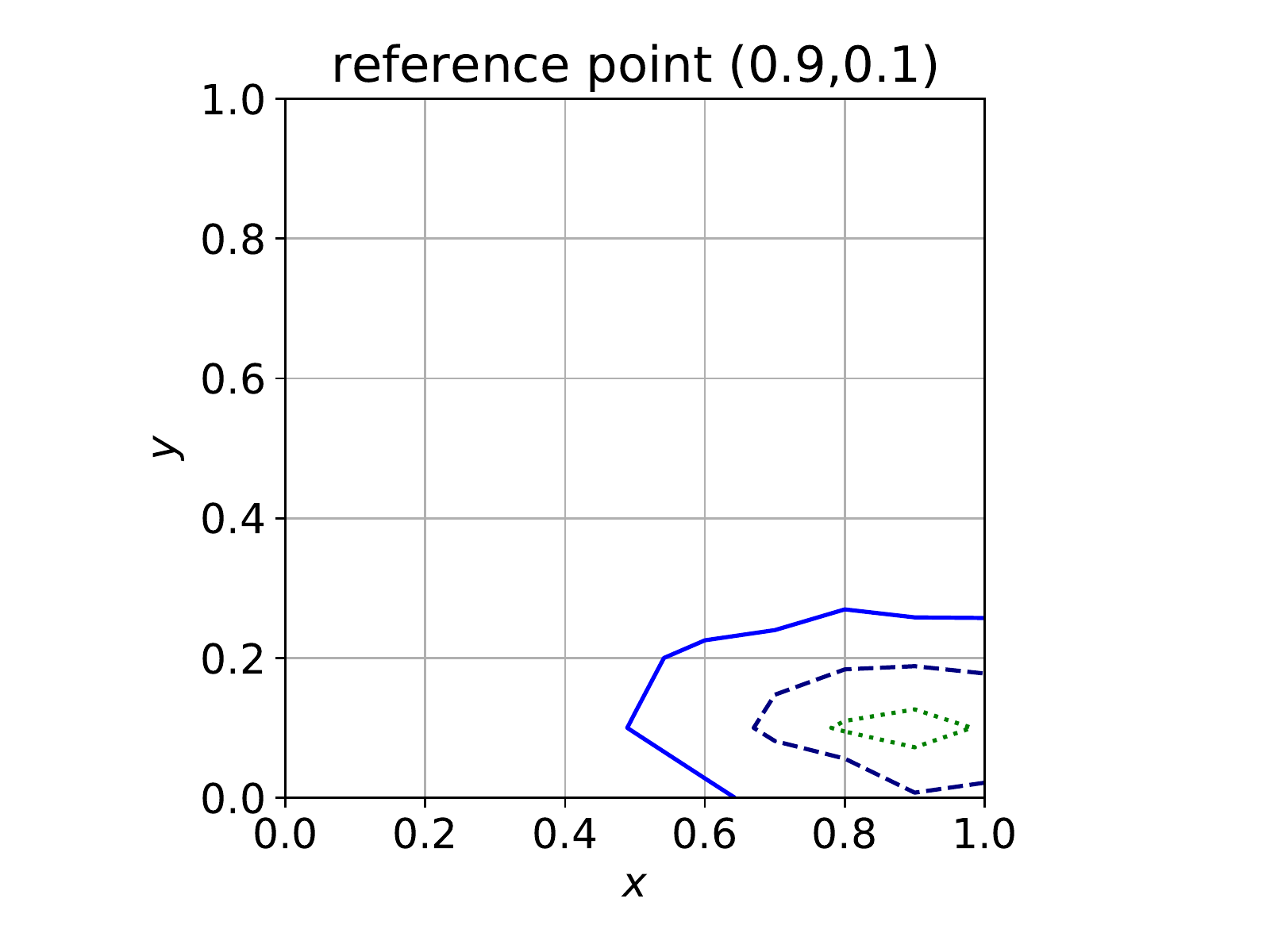}
  \includegraphics[width=0.48\hsize]{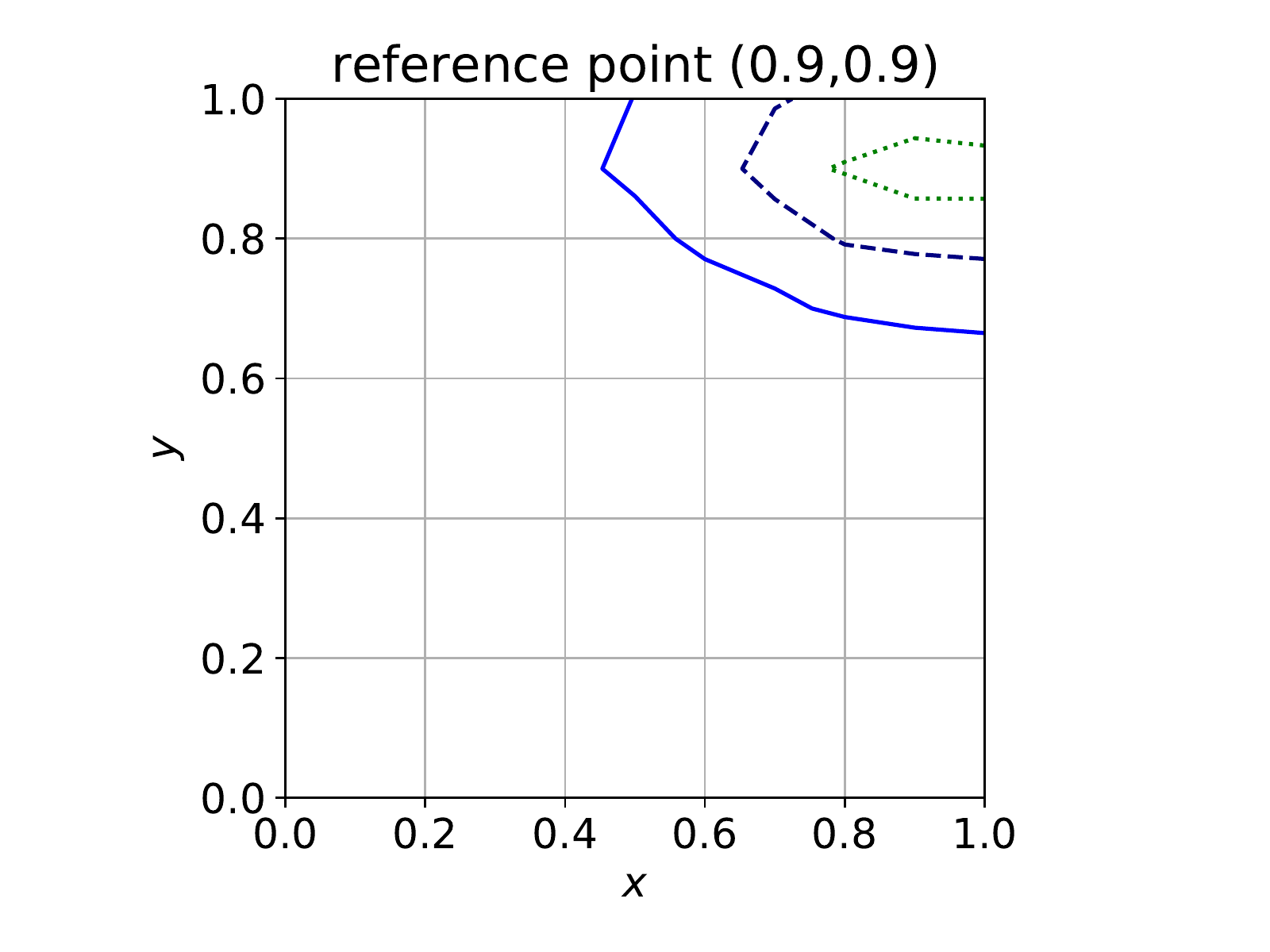}
  \caption{The expected $95\, \%$ C.L. constraints on the $(x,y)$
    plane for the case of the sample point 1.  The reference point is
    taken to be $(0.5,0.5)$ (top left), $(0.1,0.9)$ (top right),
    $(0.9,0.1)$ (bottom left), and $(0.9,0.9)$ (bottom right).  The
    green dotted, navy dashed and blue solid contours are for the integrated luminosity of
    $10$, $3$, and $1\, {\rm ab}^{-1}$, respectively.}
  \label{fig:xy_6}
\end{figure}

\begin{figure}
  \centering
  \includegraphics[width=0.48\hsize]{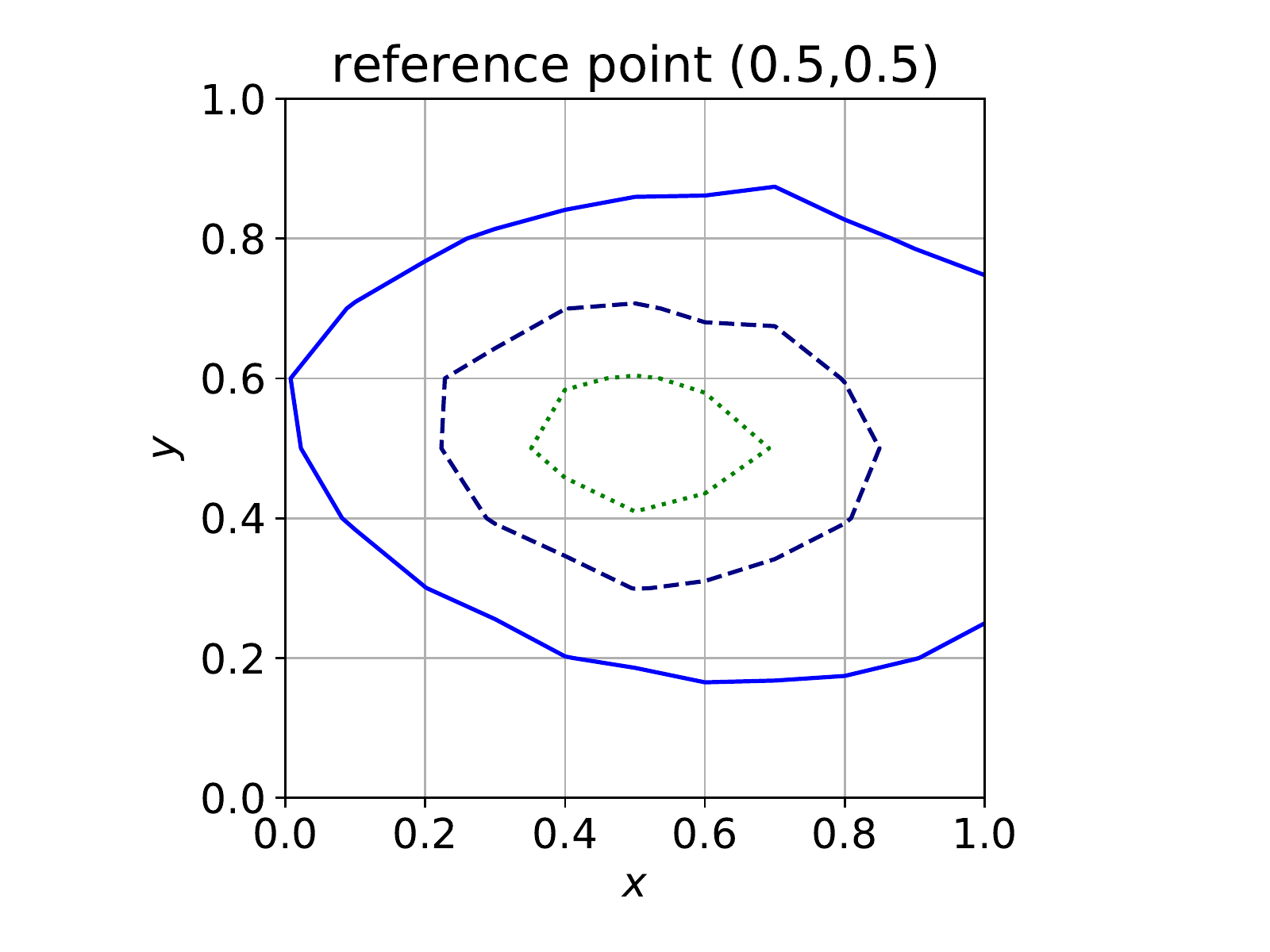}
  \includegraphics[width=0.48\hsize]{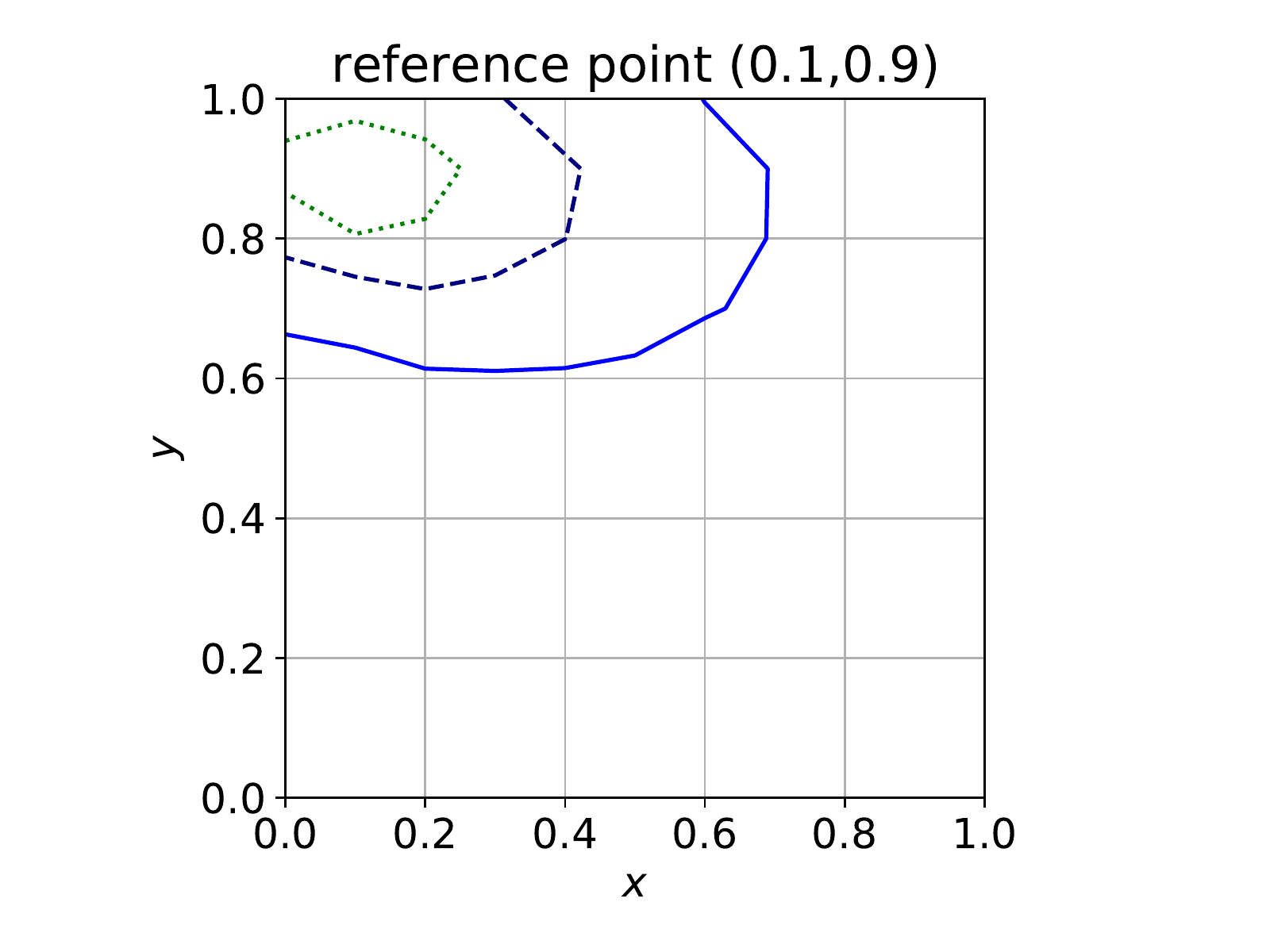}
  \includegraphics[width=0.48\hsize]{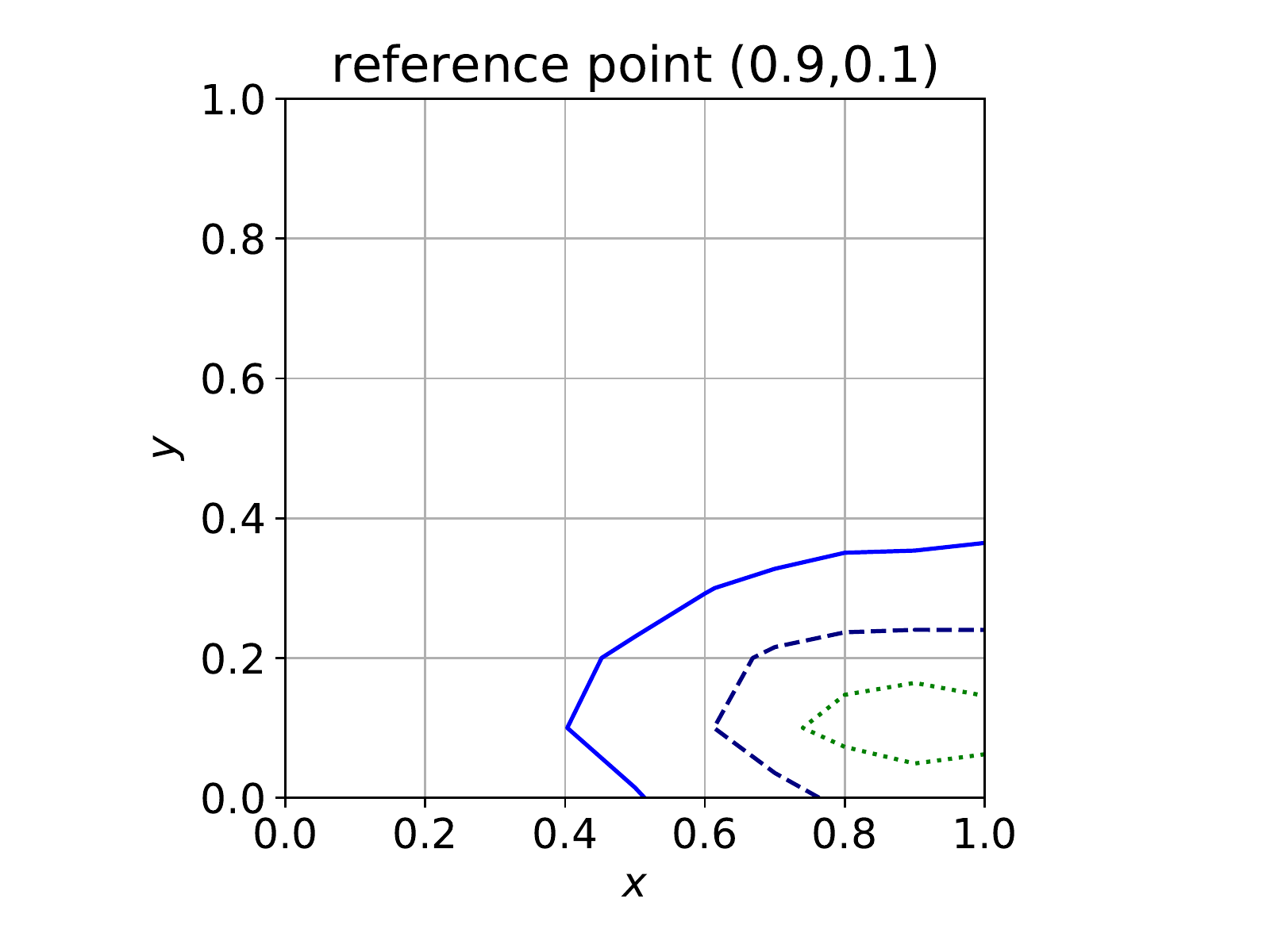}
  \includegraphics[width=0.48\hsize]{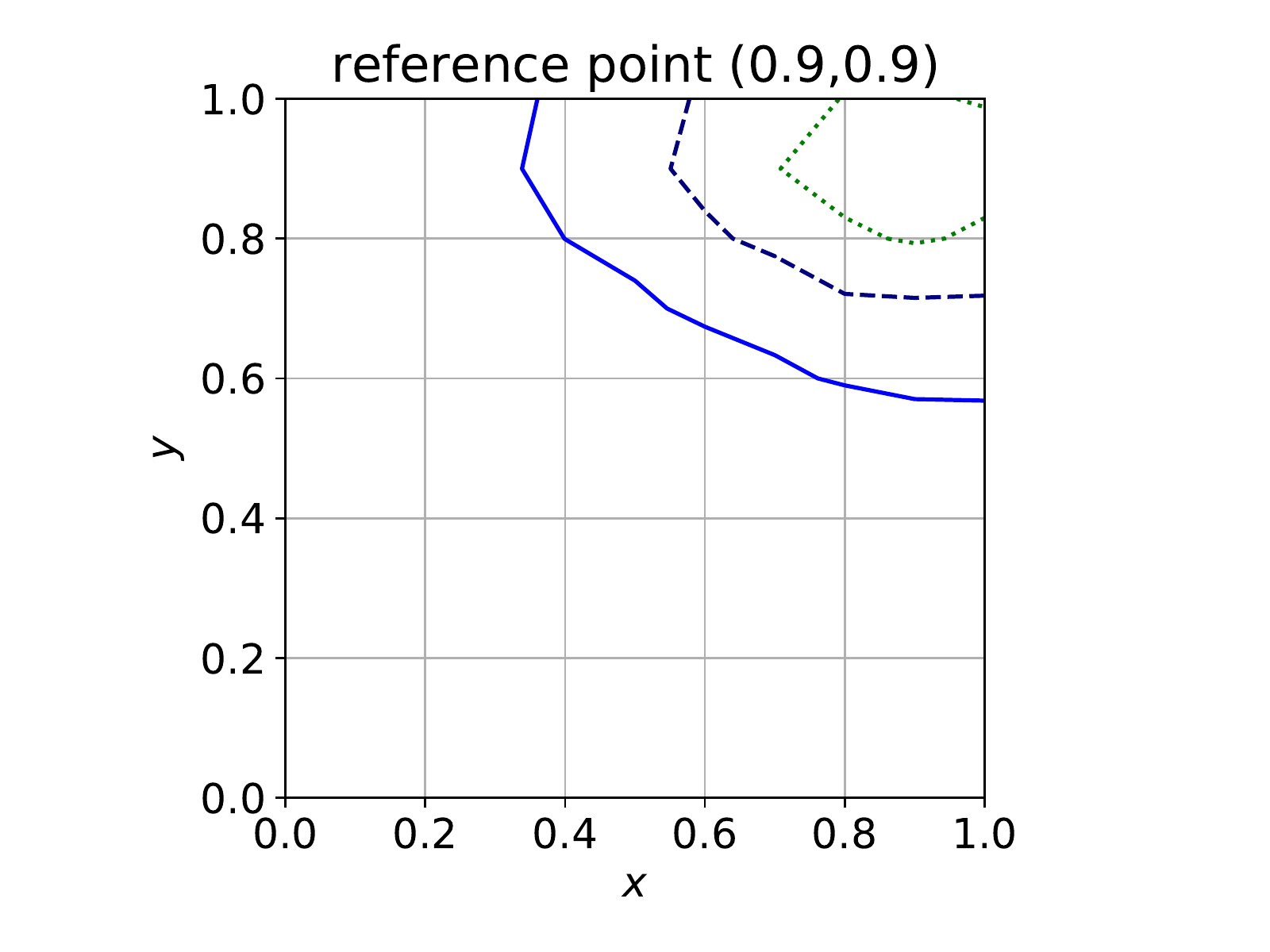}
  \caption{The expected $95\, \%$ C.L. constraints on the $(x,y)$
    plane for the case of the sample point 2.  The reference point is
    taken to be $(0.5,0.5)$ (top left), $(0.1,0.9)$ (top right),
    $(0.9,0.1)$ (bottom left), and $(0.9,0.9)$ (bottom right).  The
    green dotted, navy dashed and blue solid contours are for the integrated luminosity of
    $10$, $3$, and $1\, {\rm ab}^{-1}$, respectively.}
  \label{fig:xy_7}
\end{figure}

In Figs.~\ref{fig:xy_6} and \ref{fig:xy_7}, we show expected $95\%$
C.L. constraints on the $(x,y)$ plane for the sample points 1 and 2,
respectively, taking the integrated luminosity of $1$, $3$, and
$10\ {\rm ab}^{-1}$.  In the analysis, we take the reference points with
$(x_0,y_0)=(0.5,0.5)$, $(0.1,0.9)$, $(0.9,0.1)$ and $(0.9,0.9)$.  We can see
that the analysis of our proposal can give information about the
partial branching ratios of the gluino for both sample points 1 and 2.
The figures indicate that the accuracy of the $y$ determination
is better than that of $x$.  This is because, for the reference points
we used, the number of $b$-tagged jets is larger than that of leptons
and boosted $W$-jets.  Note that the expected accuracy of the $x$
determination can be better if more leptons can be used for the
analysis.  For example, the statistics can be improved by lowering the
$p_T$ cut for the leptons.  Currently, leptons with $p_T>200\, {\rm
  GeV}$ are used for the analysis; we have checked that, if the $p_T$
cut for the leptons can be lowered, the sensitivity to the $x$
parameter becomes better.  However, low $p_T$ leptons may be produced
by the initial state radiations which we do not simulate in our
analysis, so we do not pursue this direction.

Now we comment on possible sources of systematic errors that have not
been considered so far.  The analysis of our proposal relies on the
assumption that, once the model parameters (in the present case, $x$
and $y$, as well as the gaugino masses and the Wino lifetime) are
fixed, reliable calculations of the numbers of events in the signal
regions can be performed.  This may be the case in particular at the
time when the FCC-hh experiment will start.  The calculations of the
numbers of events, however, are likely to be affected by uncertainties
in the model, beam, and detector parameters.  Here, we perform a
simple estimation of the systematic errors by assuming that the
systematic uncertainties in the number of events to be of $\sim
10\ \%$, for example.  (More precise estimation of the uncertainty in
the determination of the model parameters $x$ and $y$ is beyond the
scope of this paper because it requires a detailed understanding of
the sources of systematic errors at the time of the FCC-hh, which is
currently quite uncertain.)  By varying the number of events in the
trial points $N_i^{(x,y)}$ in Eq.~\eqref{eq:chi2} by $\sim 10\ \%$
while fixing $N_i^{(x_0,y_0)}$,
we found that the changes of the upper and lower bounds on the $x$ are
$\sim 10\ \%$ and are smaller than the error in the $x$ determination
shown in Figs.~\ref{fig:xy_6} and \ref{fig:xy_7}.  On the contrary,
the uncertainty in the total number of events does not affect so much
the $y$ determination because, as indicated in
Fig.~\ref{fig:eventnumbers}, the number of signal event is insensitive
to the $y$-parameter.  One of the sources of the uncertainty is the
error in the gluino mass.  Analyzing the invariant mass distribution
of the decay products of the gluino, the gluino mass can be determined
with an accuracy of a few \% \cite{Asai:2019wst}, which results in
$\sim 20 \%$ uncertainty of the gluino production cross section.  The
uncertainties due to the luminosity and the parton distribution
function may be of the same size.  Another possible source of the
systematic error is the Wino lifetime.  Theoretically the Wino
lifetime can be calculated with an accuracy of a few \%
\cite{Ibe:2012sx}, while the experimental measurement of the Wino
lifetime is possible at the FCC-hh with the accuracy of $\sim 14\ \%$
by using the flight length distribution \cite{Chigusa:2019zae}.  We
checked that, if the Wino lifetime has an error of $5-10\ \%$, the
uncertainty of the total number of events is $\sim 10-20\ \%$.  Thus,
we expect that the uncertainties in the number of events can be
controlled to be $O(10)\ \%$, and that the systematic errors in the
$x$ and $y$ determinations can be smaller than the statistical ones.

\section{Implication}
\label{sec:implication}
\setcounter{equation}{0}

So far, we have seen that we can determine the partial branching
ratios of the gluino at the FCC-hh with certain accuracies if the gluino is
within the kinematical reach.  One important implication is that the
determination of the partial branching ratios can give us information
about the mass spectrum of squarks.

\begin{figure}
  \centering
  \includegraphics[width=0.48\hsize]{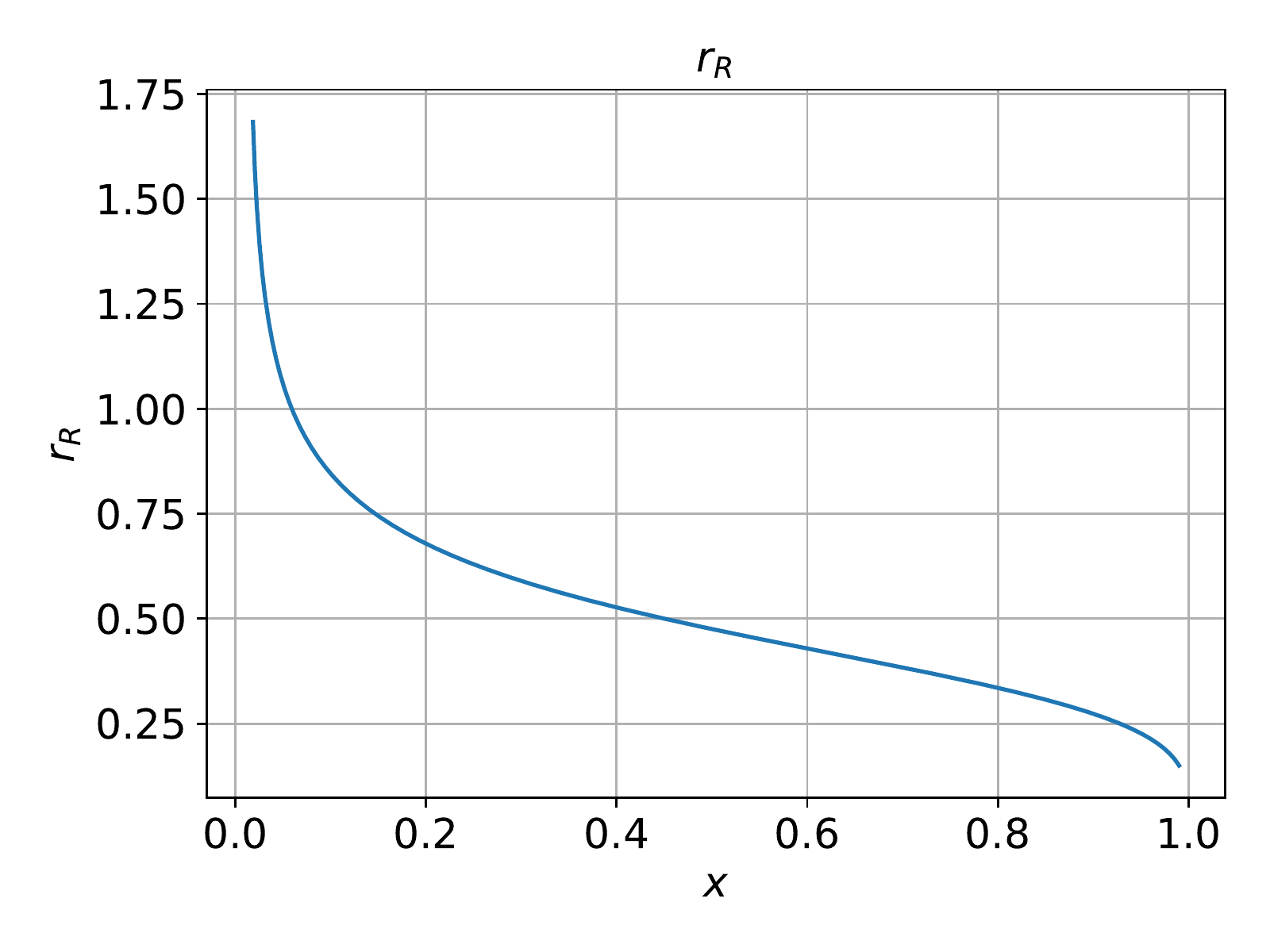}
  \includegraphics[width=0.48\hsize]{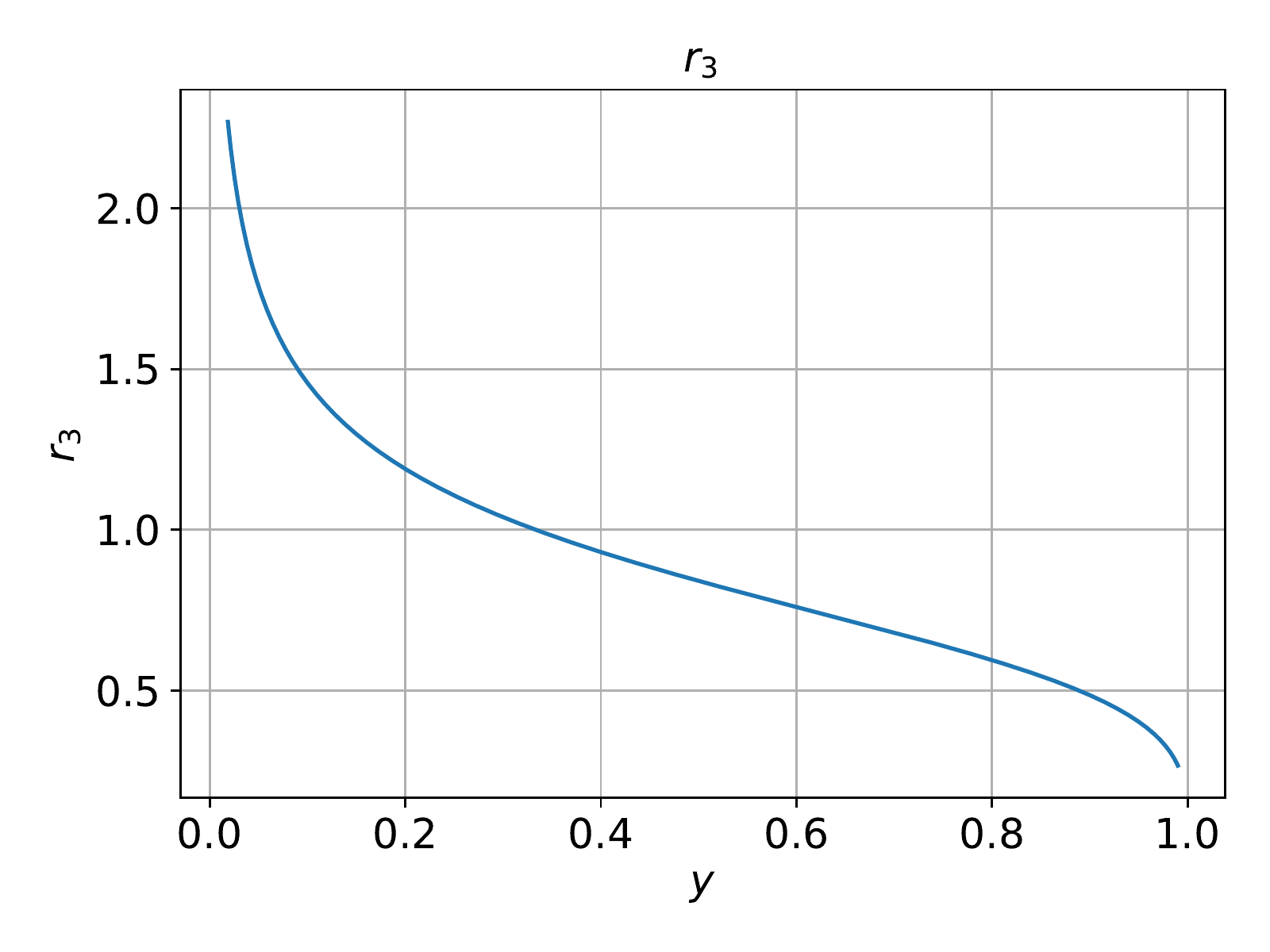}
  \caption{$r_R$ (left) and $r_3$ (right) as functions of $x$ and $y$,
    respectively, for the sample point 1.  The behavior of $r_R$ for
    the sample point 2 looks almost the same.}
  \label{fig:function}
\end{figure}

To see this, we perform a simplified analysis.  The branching ratios
in Eqs.\ \eqref{G2B} -- \eqref{G2Wpm} are realized when the squark
masses have the following relations parameterized by $r_R$ and $r_3$
(see Eqs.\ \eqref{eq:gluino_decay_Bqq} --
\eqref{eq:gluino_decay_Wpm}):
\begin{align}
  m_{\tilde{d}_i} &= \frac{1}{\sqrt{2}} m_{\tilde{u}_i} \equiv r_R m_{\tilde{Q}_i},
  \\
  \frac{m_{\tilde{Q}_3}}{m_{\tilde{Q}_{1,2}}}
  &=
  \frac{m_{\tilde{u}_{3}}}{m_{\tilde{u}_{1,2}}}
  =
  \frac{m_{\tilde{d}_{3}}}{m_{\tilde{d}_{1,2}}}
  \equiv r_3.
\end{align}
Then, the parameters $x$ and $y$ are related to
the parameters $r_R$ and $r_3$ as:
\begin{align}
  x&= \frac{ \kappa (1+4 r_R^{-4})}{27 + \kappa(1+4 r_R^{-4})},
\end{align}
with
\begin{align}
  \kappa = \tan^2\theta_W
  \frac{f(m_{\tilde{B}}/m_{\tilde{g}})}{ f(m_{\tilde{W}}/m_{\tilde{g}})},
\end{align}
and
\begin{align}
y &= \frac{1}{2r_3^4+1}.
\end{align}
Fig.~\ref{fig:function} shows the shapes of $r_R$ and $r_3$. They are
flat when the squark masses are fairly degenerate.
Thus, in such a parameter region, the determinations of $x$ and $y$
parameters can provide lower and upper bounds on squark mass ratios.
On the contrary, if the squark masses are hierarchical, we can obtain lower or upper bounds on the mass ratios.

\begin{figure}
  \centering
  \includegraphics[width=0.48\hsize]{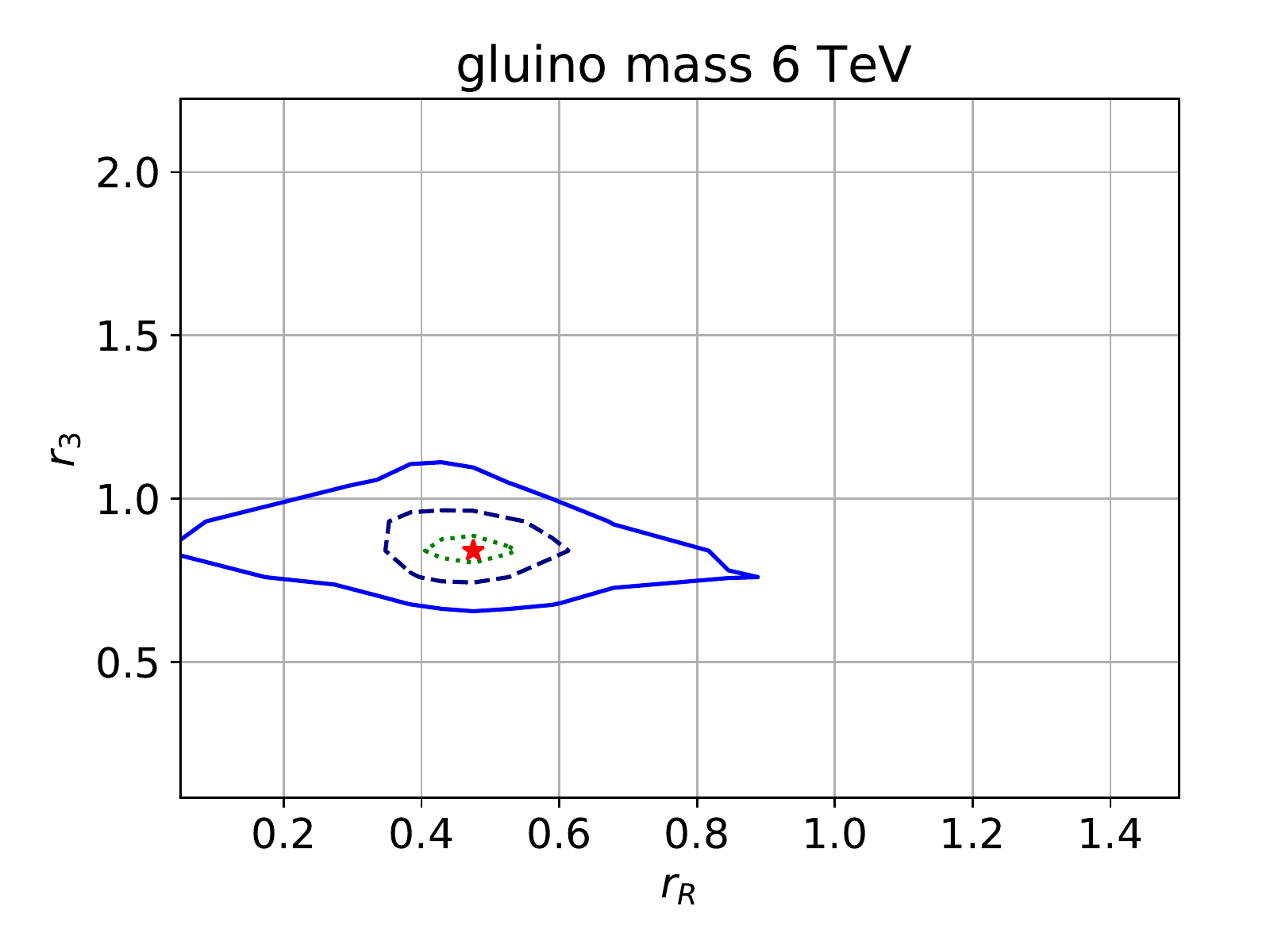}
  \includegraphics[width=0.48\hsize]{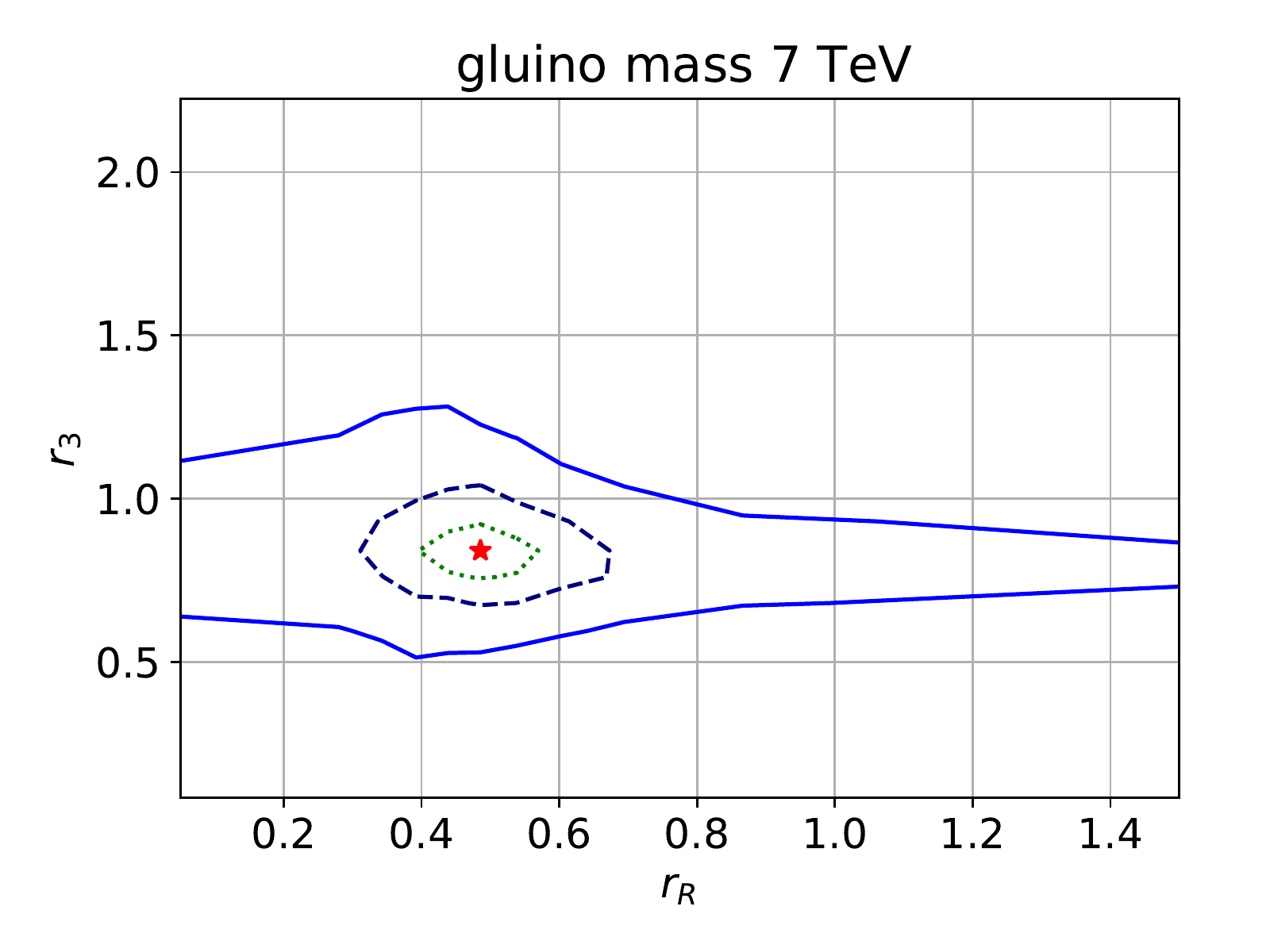}
  \caption{The $95\%$ C.L. constraints on the $(r_R,r_3)$ plane for
    the sample point 1 (left) and 2 (right).  The green dotted, navy dashed, and
    blue solid contours are for the integrated luminosity of $10$, $3$, and
    $1\ {\rm ab}^{-1}$, respectively.  The reference point corresponds
    to $(x,y)=(0.5,0.5)$ and is indicated by the red star on the
    figure.}
  \label{fig:squark_mass}
\end{figure}

In order to see how well the $r_R$ and $r_3$ parameters are
determined, we convert the constraint on the $(x,y)$ plane obtained in
the previous section to the constraint on the $(r_R,r_3)$ plane 
for the sample points 1 and 2.
 (The parameter $\kappa$ is $\kappa\simeq 0.34$ and $0.37$,
respectively.)  Fig.~\ref{fig:squark_mass} shows the contour of $95\%$
C.L.  constraint on the $(r_R,r_3)$ plane, adopting the reference
point of $(x_0,y_0)=(0.5,0.5)$.  We can see that, for the reference
point with $r_3$ and $r_R$ being both $\sim 1$, the analysis of our
proposal can determine the mass ratios of squarks.

\section{Summary}
\label{sec:summary}
\setcounter{equation}{0}

In this paper, we have discussed the possibility of studying the decay
properties of the gluino at future circular $pp$ collider with the
center of mass energy of $\sim 100\ {\rm TeV}$ (dubbed as FCC-hh).  In
the pure gravity mediation model, in which squarks are much heavier
than the gauginos, the gluino can decay as
$\tilde{g}\rightarrow\tilde{W}q\bar{q}'$ and
$\tilde{g}\rightarrow\tilde{B}q\bar{q}'$.  The gaugino in the final
state, as well as the flavors of the daughter quarks, are highly model
dependent; they depend on the mass spectrum of squarks.  We have shown
that, with the study of the number of leptons, boosted $W$-jets, and
$b$-tagged jets, FCC-hh may determine the partial branching ratios of
the gluino.  We may understand the gaugino species from the decay of
the gluino by studying the numbers of leptons and boosted $W$-jets,
while the quark flavors in the final state may be understood by
counting the number of $b$-tagged jets.  The decay properties of the
gluino are sensitive to the squark masses.  We have demonstrated that,
with the measurement of the gluino partial branching ratios, FCC-hh
can provide information about the squark mass spectrum even if squarks
are out of the kinematical reach.

\section*{Acknowledgments}

This work was supported by JSPS KAKENHI Grant Numbers
20J00046[SC], 19H05810 [KH], 19H05802 [KH], and 20H01897 [KH],
16H06490 [TM], 18K03608 [TM].
SC was supported by the Director, Office of Science, Office of High Energy Physics of the U.S. Department of Energy under the Contract No. DE-AC02-05CH1123.


\bibliographystyle{jhep}
\bibliography{ref}



\end{document}